# High-speed metamagnetic resistive switching of FeRh through Joule heating

Nicholas A. Blumenschein[1*], Gregory M. Stephen[1], Cory D. Cress[2], Samuel W. LaGasse[2], Aubrey T. Hanbicki[1], Steven P. Bennett[3], and Adam L. Friedman[1*]

[1] *Laboratory for Physical Sciences, 8050 Greenmead Dr., College Park, MD, 20740, USA*

[2] *Electronics Science and Technology Division, United States Naval Research Laboratory, 4555 Overlook Ave., SW, Washington, DC, 20375, USA*

[3] *Materials Science and Technology Division United States Naval Research Laboratory, 4555 Overlook Ave., SW, Washington, DC, 20375, USA*

[*] nblumenschein@lps.umd.edu, afriedman@lps.umd.edu

Due to its proximity to room temperature and demonstrated high degree of temperature tunability, FeRh's metamagnetic ordering transition is attractive for novel high-performance computing devices seeking to use magnetism as the state variable. We demonstrate electrical control of the transition via Joule heating in FeRh wires. Finite element simulations based on abrupt state transition within each domain result in a globally smooth transition that agrees with the experimental findings and provides insight into the thermodynamics involved. We measure a 150 K decrease in transition temperature with currents up to 60 mA, limited only by the dimensions of the device. The sizeable shift in transition temperature scales with current density and wire length, suggesting the absolute resistance and heat dissipation of the substrate are also important. The FeRh phase change is evaluated by pulsed I-V using a variety of bias conditions. We demonstrate high speed (~ ns) memristor-like behavior and report device performance parameters such as switching speed and power consumption that compare favorably with state-of-the-art phase change memristive technologies.

## I. INTRODUCTION

Magnetic materials are crucial components of memory devices because of their inherent non-volatility, radiation hardness, and ease of control [1–4]. Using magnetization as a state variable has several advantages over charge-based devices [5]. For one, the resonant frequencies within magnetic materials are at least an order of magnitude faster than existing DRAM technology: gigahertz for ferromagnets, and terahertz for antiferromagnets [6]. Presently, magnetic memory relies on manipulating the magnetic moment direction [7]. Although this scheme is non-volatile, it can be difficult to distinguish between on/off states in these devices. Also, future MRAM technology relies on switching magnetic tunnel junctions with spin transfer torque, which requires high currents to switch the free magnetic layer. This can degrade the tunnel barrier rather quickly, rendering the device inoperable [8,9].

An alternative approach to employ magnetization-based devices is to toggle the magnetization state itself by switching between a ferromagnetic (FM) and antiferromagnetic (AFM) phase, giving a clear on/off state in a phase-change device. Phase-change memory (PCM) devices are typically operated based on the resistivity contrast of insulating amorphous and conductive crystalline phases [10,11]. Indeed, PCM devices capable of faster write times and higher endurance than traditional NAND memory are within reach [11]. Memory and logic elements with bistable states are another unique possibility if the switching excitation is kept near the activation energy of the phase transition. Such bistable transistors are envisioned for neuromorphic computing architecture applications, although existing devices are slow and lithographically complicated [12–14]. FeRh provides an ideal platform for fast, lithographically simple phase-change memories because of its novel AFM to FM transition that is accompanied by volumetric expansion of the CsCl-type crystal lattice and a significant change in resistivity [15–17].

Furthermore, the critical temperature ($T_{Cr}$) at which the AFM-FM transition occurs is near room temperature and can be tuned using substitutional doping [18], strain [13,19–29], and patterning [26,30,31]. Because the transition is temperature-dependent, device manipulation through Joule heating is also possible [32]. An electric current flowing through an FeRh wire can heat the material above $T_{Cr}$, inducing the AFM-FM transition [31,33]. Previous reports on FeRh suggest that the transition is very rapid, occurring on a time scale of ≤ 500 fs [34]. This could lead to a new class of PCM devices operating at THz frequencies perfectly suited to neuromorphic computing applications [35].

In this work we demonstrate fast resistive switching of FeRh wires through Joule heating. We find the shift in AFM-FM transition temperature scales with both current density and wire geometry. We use pulsed I-V measurements to investigate the dynamic Joule heating effects and resulting power switching losses accompanying the AFM-FM transition. We obtain device switching speeds of about 300 ns, a value limited by our measurement equipment. We perform finite element method-based simulations to confirm the explanation for the observed behavior and provide further insight into the heat-induced transition.

**II. EXPERIMENTAL DETAILS**

FeRh films were grown on MgO substrates by sputter deposition with thicknesses of 35 and 200 nm. A 500 nm $SiO_2$ layer was deposited using plasma-enhanced chemical vapor deposition (PECVD) as a hard mask for FeRh ion milling. Standard photolithography was used to define 10 × 100 μm Hall bars and wires of varying widths and lengths. Electron beam lithography (EBL) was used to define the wire geometry in poly-methyl methacrylate (PMMA) followed by metal deposition of a 20 nm Cr hard mask on the $SiO_2$ surface. The $SiO_2$ was then etched by inductively coupled plasma (ICP) reactive ion etching (RIE) with 10 sccm $CHF_3$ and 15 sccm Ar with a chamber pressure of 30 mTorr, 30 W RF power, and 600 W ICP power, defining a hard mask for the ion milling process. The Cr mask was removed by the ion milling process and any remaining $SiO_2$ over the FeRh wires is removed by a second ICP/RIE step. Finally, EBL was used to define bond-pads in PMMA followed by electron beam evaporation of Ti/Au (10/50 nm) and metal lift-off in acetone. Fabricated FeRh wires are shown in **Figure 1(a, b)**.

The FeRh crystalline quality was evaluated using X-ray diffraction and high-resolution transmission electron microscopy, and the details for representative films have been previously reported elsewhere [36]. Resistance measurements were taken in a closed-cycle cryogenic probe station using a parameter analyzer. All measurement data was gathered while the sample was under a vacuum of $<1 \times 10^{-4}$ Torr. COMSOL Multiphysics software was used for finite element method simulation and analysis of the electro-thermal transport properties of FeRh and surrounding materials.

**III. RESULTS**

FeRh films were processed into Hall bars and two-terminal wire devices with varying dimensions. Representative two-terminal devices are shown in **Figure 1(a)**, with the enhanced image containing an FeRh device with a wire thickness, width, and length of 35 nm, 1 μm, and 100 μm, respectively. **Figure 1(b)** shows a three-dimensional optical image of the two-terminal FeRh device as measured using a confocal laser microscope. **Figure 1(c)** shows a behavioral schematic of an FeRh Hall bar with film thickness of 200 nm. Here, the resistance is plotted as a function of ambient temperature while applying a constant 25 mA current (2.5 MA $cm^{-2}$). The background shading colors denote the three different temperature regimes at which the FeRh is AFM (blue), FM (red), and in transition (white). Upon heating the device (red curve), we see the FeRh begin the AFM-FM transition once the temperature exceeds $T_{Cr}$ (T > 355 K). The phase transition is accompanied by a decreasing resistance, persisting until T > 420 K where the FeRh has fully transitioned into the FM phase. A similar-but-opposite effect occurs when cooling the device (blue curve; FM-AFM transition for 345K < T < 410 K, AFM transition when T < 345 K). In **Figure 1(d),** we display the resistivity ($\rho_{xx}$) of a 200-nm-thick FeRh Hall bar (inset) as a function of temperature for current densities ranging from J = 1 MA cm$^{-2}$ to 5 MA cm$^{-2}$ (10 mA to 50 mA, respectively), measured while sweeping the sample temperature at a rate of ±1 K min$^{-1}$. The AFM-FM transition, as evidenced by the abrupt change in resistivity, is observed for each applied current density regardless of ambient temperature, but shifts from 410 K at 1 MA cm$^{-2}$ to 255 K at 5 MA cm$^{-2}$. The decrease in transition temperature is accompanied by a widening of the transition range that can be explained by our model, as discussed below and in the appendix.

With the current-dependence of the AFM-FM transition established, we then investigated the transition at fixed temperatures while sweeping current in the FeRh wires. **Figure 2(a)** shows the wire resistivity as a function of current density at sample temperatures ranging from 300 K to 480 K for a 0.3 × 100 µm wire with thickness of 35 nm. At 300 K the FeRh remains in the AFM phase as the induced Joule heating is insufficient to elevate the wire temperature to $T_{Cr}$. At 320 K the AFM-FM transition begins at a high current density of ~ 3.3×10$^7$ A cm$^{-2}$. At 340 – 360 K we observe a larger region of the AFM-FM transition, but the current density is insufficient to force the FM transition. At 380 – 400 K the wire fully transitions into the FM phase, then when the current density is decreased the FM-AFM transition begins as the wire is cooled. At 420 – 440 K the wire begins in the AFM phase, then transitions into the FM phase with an increasing current density. However, the high ambient temperature prevents the FeRh from transitioning back into the AFM phase when reducing the current density to zero. At 460 – 480 K the wire begins in the FM phase, as the AFM-FM transition was fully driven by ambient temperature. The shift in transition temperature, measured as the maximum of the ρ-J curve where ρ begins to decrease, is shown in the inset of **Figure 2(d)**. Joule heating depends on applied power and is directly proportional to the square of the current. Therefore, the reduced current density required for the AFM-FM phase transition at higher temperatures is a clear indication of Joule heating. At an ambient temperature within the transition width, the current-induced Joule heating partially transitions the FeRh, then returns to a lower resistance as the current is reduced.

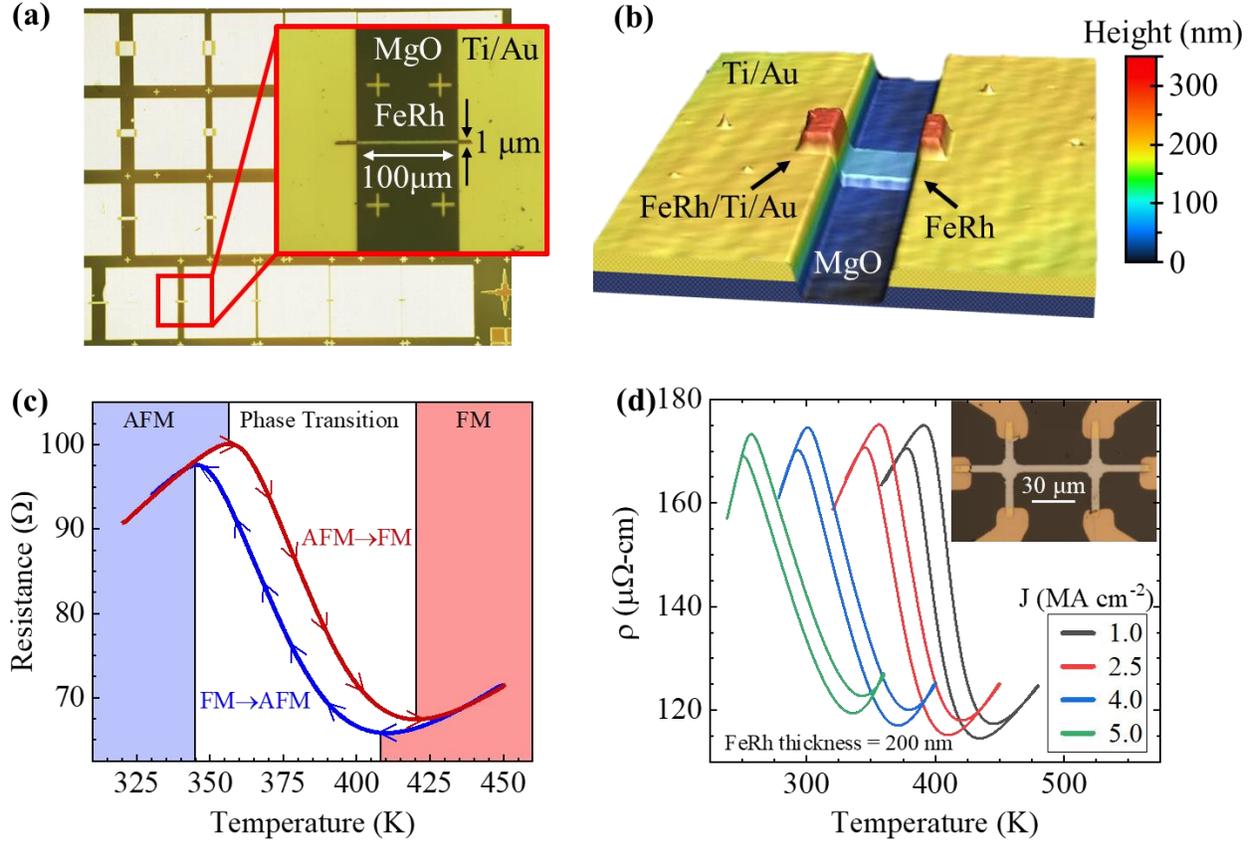

FIG. 1. (a) Top-view optical images of the fabricated two-terminal FeRh devices with 35-nm-thick wires with widths and lengths varying from 0.3 to 50 µm and 2.5 to 100 µm, respectively. The enhanced image shows an FeRh wire with width, height, and thickness of 1 µm, 100 µm, and 35 nm, respectively. (b) Three-dimensional optical image of the device showing topology. (c) Proof-of-concept FeRh wire resistance while varying the ambient temperature from 320 K to 450 K. Red and blue curves represent heating and cooling cycles, respectively. Background shading colors denote temperature regimes at which the FeRh is AFM (blue), FM (red), and in transition (white). (d) FeRh Hall bar resistivity as a function of ambient temperature while varying the current density through the device. The inset shows an image of the FeRh Hall bar with FeRh layer thickness of 200 nm.

Wire geometry dependencies were investigated to further evaluate thermal dissipation effects. These measurements were performed at 400 K. **Figure 2(b)** shows the ρ-J curves for 1 μm wide FeRh wires with lengths ranging from 2.5 μm to 100 μm and thickness of 35 nm. The resistivity scale for each curve is offset for clarity. The current density through a wire does not scale as a function of wire length. Nonetheless, here we observe an AFM-FM transition temperature that is reduced for longer wires. Applying a constant current density through wires of increasing length requires a larger applied bias. Regardless of the associated thermal profile, longer wires dissipate more power which causes the substrate temperature to increase. Similarly, as shown in the ρ-J curves of **Figure 2(c)**, the wire width also effects AFM-FM transition temperature. The data shown here were measured using 100 μm long FeRh wires with widths ranging from 0.3 μm to 50 μm. The curves are offset for clarity. The larger wires cannot cool as quickly, as a larger FeRh/substrate interfacial area results in a greater substrate heating contribution and lessened heat dissipation into the metal contacts and surrounding ambient. The 100 μm length was chosen because many of the shorter wires do not fully transition through the hysteresis region within the given current density range.

**Figure 2(d)** shows the critical power dissipation, $P_{Cr}$, and critical current density, $J_{Cr}$ (inset), as a function of temperature for two 100 μm long FeRh wires with widths of 0.3 μm (red squares) and 1.0 μm (blue circles). As shown in the inset, $J_{Cr}$ does not decrease linearly with increasing temperature, consistent with a previous report [37]. However, a linear dependence is observed when plotting $P_{Cr}$ vs. temperature. The critical power dissipation is calculated by

$$P_{Cr} = I_{Cr}^2 R_{Cr} \triangleq \alpha \Delta T, \qquad (1)$$

which is expressed in terms of critical current ($I_{Cr}$) and critical resistance ($R_{Cr}$) at which the FeRh AFM-FM transition begins. Once the FeRh wire has reached thermal equilibrium, we define $P_{Cr} = \alpha \Delta T$, where $\alpha$ [W K$^{-1}$] describes heat dissipation from the FeRh wire and $\Delta T$ is the temperature increase caused by Joule heating. The linear dependence of $P_{Cr}$ on $T$ indicates that the AFM-FM phase transition is directly caused by Joule heating. Notably, an identical $T_{Cr}$ of 415 K is observed for both wire dimensions when extrapolating $P_{Cr}$ to the x-axis. The slope of the two $P_{Cr}$ curves vary because of $\alpha$, which is found to be -1.09 W K$^{-1}$ and -1.66 W K$^{-1}$ for wire widths of 0.3 μm and 1 μm, respectively. The observed increase of $\alpha$ amplitude for the larger wire width is a function of the FeRh/metal interfacial area, consistent with previous reports [37].

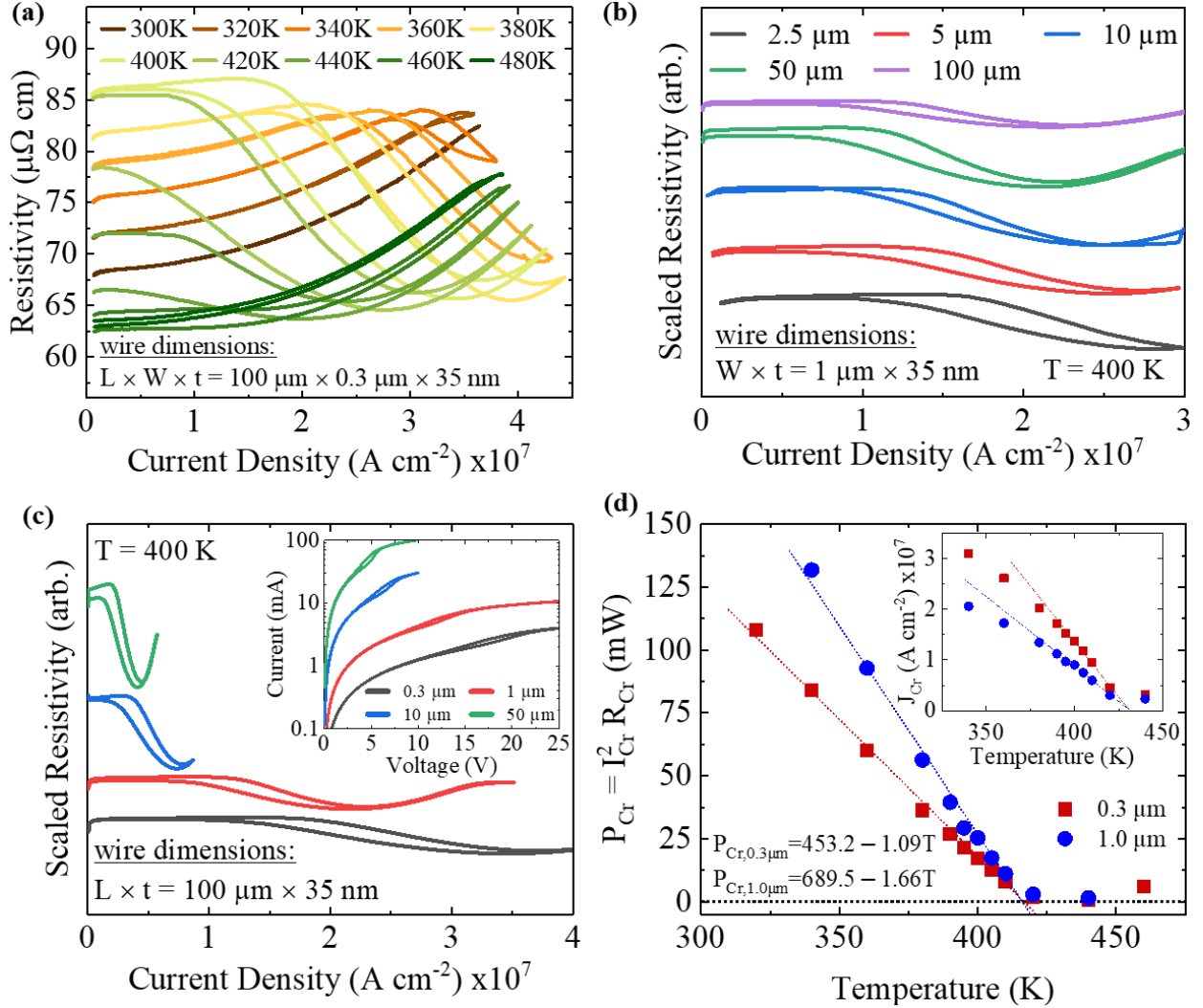

FIG. 2. FeRh magnetic phase transition analysis by DC transport characterization to investigate temperature and geometrical dependencies. (a) FeRh resistivity as a function of current density for ambient temperatures ranging from 300 K to 480 K. The experimental data shown here were obtained by measuring a FeRh wire with width, length, and thickness of 0.3 µm, 100 µm, and 35 nm, respectively. The critical current density, $J_{Cr}$, decreased significantly as the ambient temperature was increased. Resistivity was then measured as a function of current density for a series of wires with varying (b) lengths and (c) widths. The wire-length-dependent ρ-J measurements were performed using a 1-µm-wide wire, while the wire-width-dependent ρ-J measurements were performed using a 100-µm-long wire. The inset of (c) shows the measured I-V characteristic for the wires of varying width. (d) Critical power dissipation is shown as a function of ambient temperature for 100-µm-long wires with widths of 0.3 µm (red squares) and 1 µm (blue circles). The inset plot in (d) shows the critical current density for the same two wires. Joule heating phenomenon was confirmed by fitting the $P_{Cr}$-T data, during which an identical $T_{Cr}$ of 415 K was extracted from the two wires with different geometries.

Above and below the metamagnetic transition, the electrical transport properties follow that of typical metals, where increased phonon scattering degrades conductance leading to a positive temperature coefficient of resistance [38]. Moreover, the magnitude of the temperature coefficient of resistance is approximately equal in both the AFM and FM states. The absolute value of the resistivity, not its temperature dependence, differs. Near the metamagnetic transition temperature, microscopic domains of FeRh in the AFM state begin to transition into the FM state, simultaneously changing the overall resistance of the conductor. The precise temperature at which each domain changes is driven by defects, strain, the state of neighboring domains, *etc.*, creating a distribution of temperatures over which the full transition occurs [39]. Therefore, for a single domain, the change in resistance with temperature would appear abrupt and square-like, while the distribution of transition temperatures in a multi-domain wire would induce a gradual change. The origin of these thermal mechanisms, as well as the observed Joule heating behavior, are elucidated *via* finite element simulations. Heat transfer physics were modeled using temperature-dependent thermodynamic ($C_p$, κ, density) and electrical properties. In particular, the temperature-dependent conductivity of FeRh in the AFM and FM states was defined as shown in **Figure 6(a)** (Appendix). Additional modeling details are found in the Appendix.

The simulated wire dimensions and biasing conditions (Appendix **Figure 6(b)**) closely match those of **Figure 2(a)**, and the simulation captures the primary experimental behavior, verifying the modeling approach. In particular, in the model we observe a hysteresis of approximately 10 – 15 K, a reduction in the onset of transition temperature with DC bias, and an increase in the transition width with DC bias, all of which are reflected in the measured results. In appendix **Figure 6(c)** and **6(d)** we simulate the wire length and width dependencies that were measured experimentally and shown in **Figure 2(b)** and **(c)**. These trends are consistent with the experimental data, where the ampacity significantly increases for the narrowest wire. This effect is caused by the ampacity scaling as a function of the wire linear mass density [36].

Analysis of the model parameters during a temperature sweep provides insight into the fundamental origins of the globally measured transport properties. **Figure 3(a)** represents wire resistivity as a function of substrate temperature for a current density of 5 MA cm$^{-2}$. **Figure 3(b)** shows the device regions depicted in (*i*) – (*vi*), consisting of one metal contact and half of the 100-µm-long FeRh wire. In **Figure 3(c)** we show surface maps of temperature, resistivity, and state of individually modeled FeRh domains for substrate temperatures of (*i*) 300 K, (*ii*) 365 K, (*iii*) 392 K, (*iv*) 422 K, (*v*) 412 K, and (*vi*) 380 K. In (*i*), despite a relatively low substrate temperature of 300 K, the wire temperature increases to 334 K near the center. The elevated temperature is an effect of Joule heating, and it persists until reaching the edge of the contact where heat dissipation is enhanced by the Au electrode. Once the substrate temperature is raised to 365 K (*ii*), the peak wire temperature and terminal resistivity reach 417 K and 100 μΩ-cm, respectively. This temperature corresponds with the peak wire resistance, where temperature-induced increases in resistance are offset by the accumulation of domains which have transitioned to the lower resistivity FM state. These transitioned regions are observed as green domains and white domains in the resistivity and state maps in (*ii*), and are most concentrated near the middle of the FeRh wire where the temperature is the highest. This behavior becomes more pronounced as the temperature is increased to 392 K (*iii*). The locations of the transitioned regions are random, yet generally located near the middle and away from the metal contact where the wire is at its highest localized temperature. At 422 K (*iv*), the wire reaches its minimum resistance state, where the temperature-induced increases in resistance exceed the reduction in resistance from switching domains, since nearly all domains are now in the FM state. The hysteresis in the wire is clearly evidenced on the cooling cycle as nearly all domains remain in the same state at 412 K (*v*), and despite having a 12 K cooler substrate at 380 K (*vi*) we observe resistivity and state maps that closely resembles those observed upon heating to 392 K (*iii*).

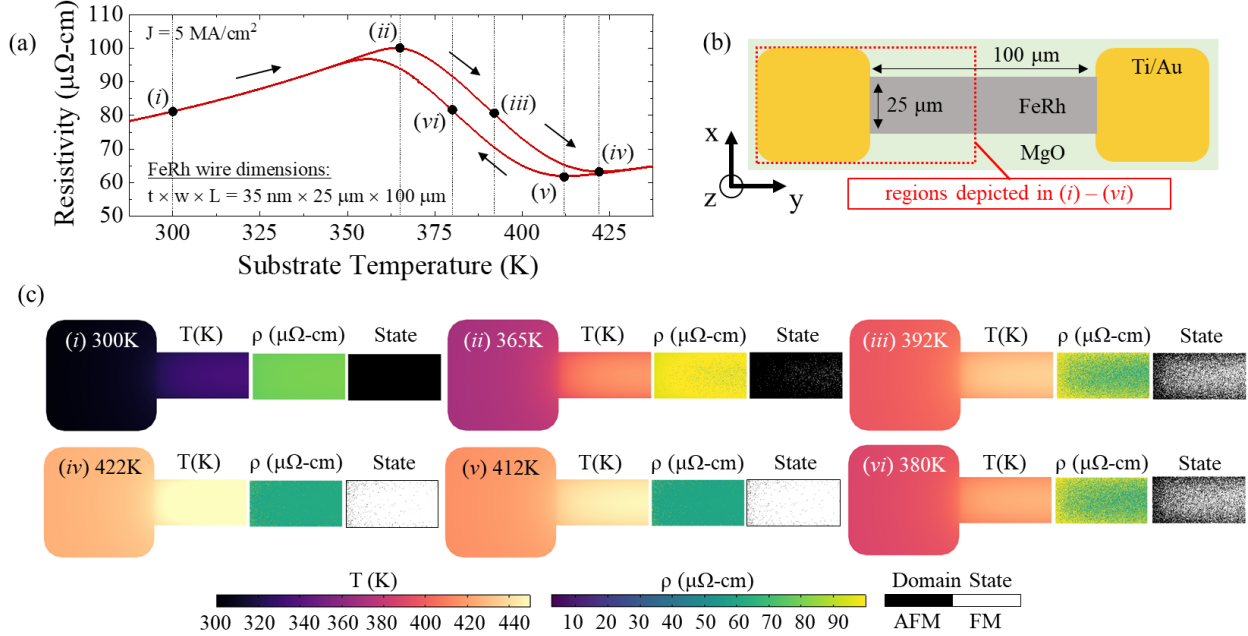

FIG. 3. (a) Representative resistance *vs.* temperature sweep for a current density of 5 MA cm$^{-2}$. Roman numeral labels indicate the specific temperatures at which the subsequent surface maps were generated. The surface maps were generated while sweeping the temperature during a heating cycle, as indicated by the data points shown on the curve. (b) A schematic diagram of the FeRh wire device including a region outlined with a red dotted line to indicate the portion of the sample which is depicted in the subsequent surface maps, including half of the FeRh wire length one metal contact. (c) Surface maps of the outlined sample area show temperature, wire resistivity, and individual FeRh domain states (AFM or FM) for substrate temperatures of (*i*) 300 K, (*ii*) 365 K, (*iii*) 392 K, (*iv*) 422 K, (*v*) 412 K, and (*vi*) 380 K. For (*v*) and (*vi*) the substrate is cooled from 500 K, all others are for the heating phase of the heat/cool cycle. The state of the domains (AFM/FM) and corresponding local resistivity correlate closely with the thermal profiles shown in the appendix (**Figure 7**).

A pulsed bias allows more precise control over self-heating effects and establishes realistic device operation parameters such as switching speeds. Pulsed I-V has been used to analyze thermal effects in a wide variety of materials, including Si [40], wide bandgap semiconductors such as GaN and $Ga_2O_3$ [41,42], and phase change materials like $Ge_3Sb_2Te_6$ [43]. Pulsed J-V measurements shown in **Figure 4(a)** were made at 400 K on a FeRh device with length and width of 100 μm and 0.3 μm, respectively. The pulse profile, shown schematically in **Figure 4(a)** includes a 5 ms pulse width (*PW*), 10 ms period, a pulsed amplitude voltage ($V_a$) varied from 0 – 30 V and a baseline voltage ($V_b$) varied from 0 – 20 V. The figure shows the J-V curves offset for visual clarity when $V_b$ = 0 V and 10 – 20 V. A DC measurement is included for comparison purposes. The inset presents the data without the offset to show how the data collapses onto a single J-V loop. The hysteresis regions of the J-V curves remain fixed along the voltage axis, but widen along the current axis as $V_b$ is increased from 10 – 20 V.

**Figure 4(b)** shows the change in resistance with current density (R-J), offset for clarity. Hysteresis in the R-J characteristic begins to develop once $V_b$ > 10 V. Although hysteresis is always seen in the DC measurements, the absence of hysteresis for low $V_b$ during pulsed operation indicates that heat is dissipating at a high enough rate to allow the FeRh to cool and transition back into the AFM phase between each pulse. This thermal dissipation effect can be manipulated by increasing the baseline voltage, evident by the evolution of the R-J hysteresis with bias voltage. When using pulses, a persistent FeRh FM state is only achieved when (1) the $V_b$ amplitude is large enough such that the FeRh can exceed $T_{Cr}$ and (2) $V_a \geq V_b$.

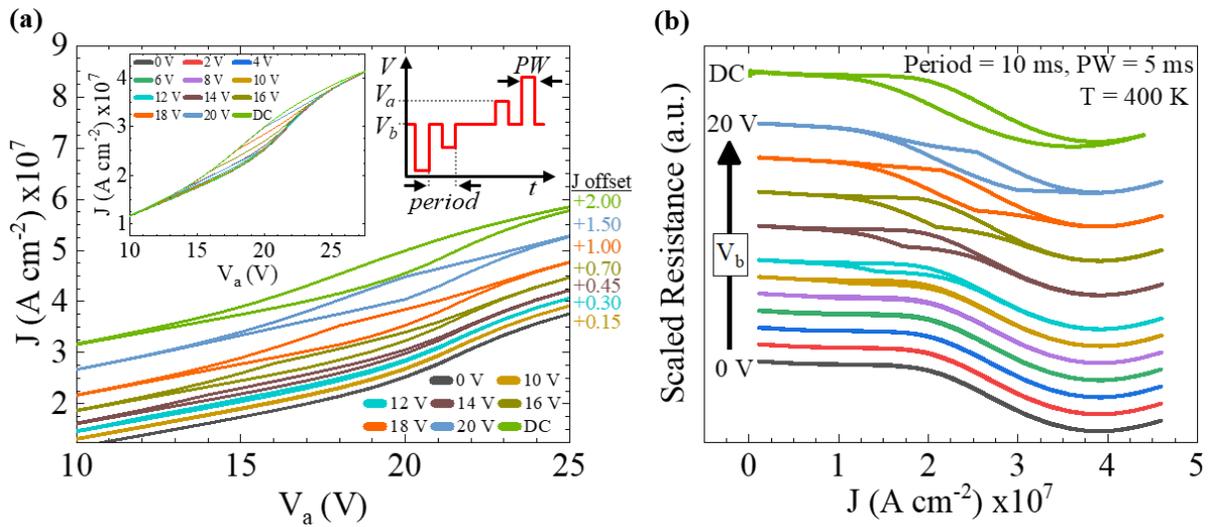

FIG. 4. A FeRh wire was characterized by pulsed J-V to investigate operating conditions which are more coherent with switching devices used in practice. The data shown here were acquired at an ambient temperature of 400 K using a FeRh wire with length, width, and thickness of 100 μm, 0.3 μm, and 35 nm, respectively. (a) J-$V_a$ and (b) R-J curves are shown while varying the base voltage amplitude. Aside from the inset plot of (a), each curve is offset for clarity and both plots contain a DC measurement for comparison purposes. Data were gathered using bias pulses of 10 ms period and 5 ms pulse width. The inset plot in (a) contains the as-measured J-V curves without an offset to show how they collapse onto a single loop. Additionally, a pulse profile schematic containing the pulse width (PW), baseline voltage ($V_b$), and amplitude voltage ($V_a$) is shown in (a).

The pulsed voltage conditions can be fine-tuned to quickly switch between the AFM and FM states. An example of this methodology is shown using the DC-IV plot of **Figure 5(a)** for FeRh wire device (width = 0.3 μm, length = 100 μm). To function as a switch, the FeRh temperature must be stabilized such that the resistance lies somewhere within the metamagnetic transition region. For example, this could be achieved using a 20 V DC bias. Now, rather than applying a constant DC bias, we can apply a pulsed bias with a 20 V baseline voltage where the FeRh settles to a temperature within the metamagnetic transition region. At this temperature, crystal domains can exist in either an AFM (high resistance) or FM phase (low resistance). The system does not need to completely switch to a pure AFM or FM phase, it just needs to have a majority of domains in either the AFM or FM phase to produce a sufficient change in resistance. When $V_a$ is set to 20 V, the AFM and FM phases have resistances of 7.4 kΩ and 6.3 kΩ, respectively. For example, the device can be switched ON (AFM-FM transition) by applying a pulsed voltage of $V_a = 30$ V then reducing $V_a$ back to 20 V to maintain state (blue arrow). Likewise, it can be switched OFF (FM-AFM transition) by applying a pulsed voltage of $V_a = 5$ V then increasing $V_a$ back to 20 V to maintain state (red arrow).

In **Figure 5(b)** we demonstrate switching functionality using the methodology described above. The transient resistance (top blue curve) was measured while applying a pulsed voltage waveform (bottom red curve). The pulse profile consists of a 20 V baseline voltage, and short 5 ms voltage pulses to either 5 V or 30 V to switch between OFF and ON states, respectively. First, the voltage was ramped from 0 to 20 V to enter the hysteresis center-region and held at 20 V for ~ 1 second, maintaining the AFM state. At $t = 1.1$ s, the voltage is then pulsed to $V_a = 30$ V for 5 ms, causing the wire temperature to increase such that the AFM-FM transition occurs and the device is switched ON. After this 5 ms pulse, $V_a$ is reduced to 20 V to maintain the FeRh temperature and hold the device in the ON state. Next, at $t = 1.9$ s, the device is switched OFF by applying a 5 ms pulse of $V_a = 5$ V, then returning to $V_a = 20$ V to maintain state. We find that $\Delta R/R_{min} = 16.4\%$ when switched into the OFF state, which is ~ 3x more resistance modulation than what has been reported for strain-based FeRh switching [19,28]. Device switching endurance and state retention were also evaluated and the results are shown in **Figure 9**.

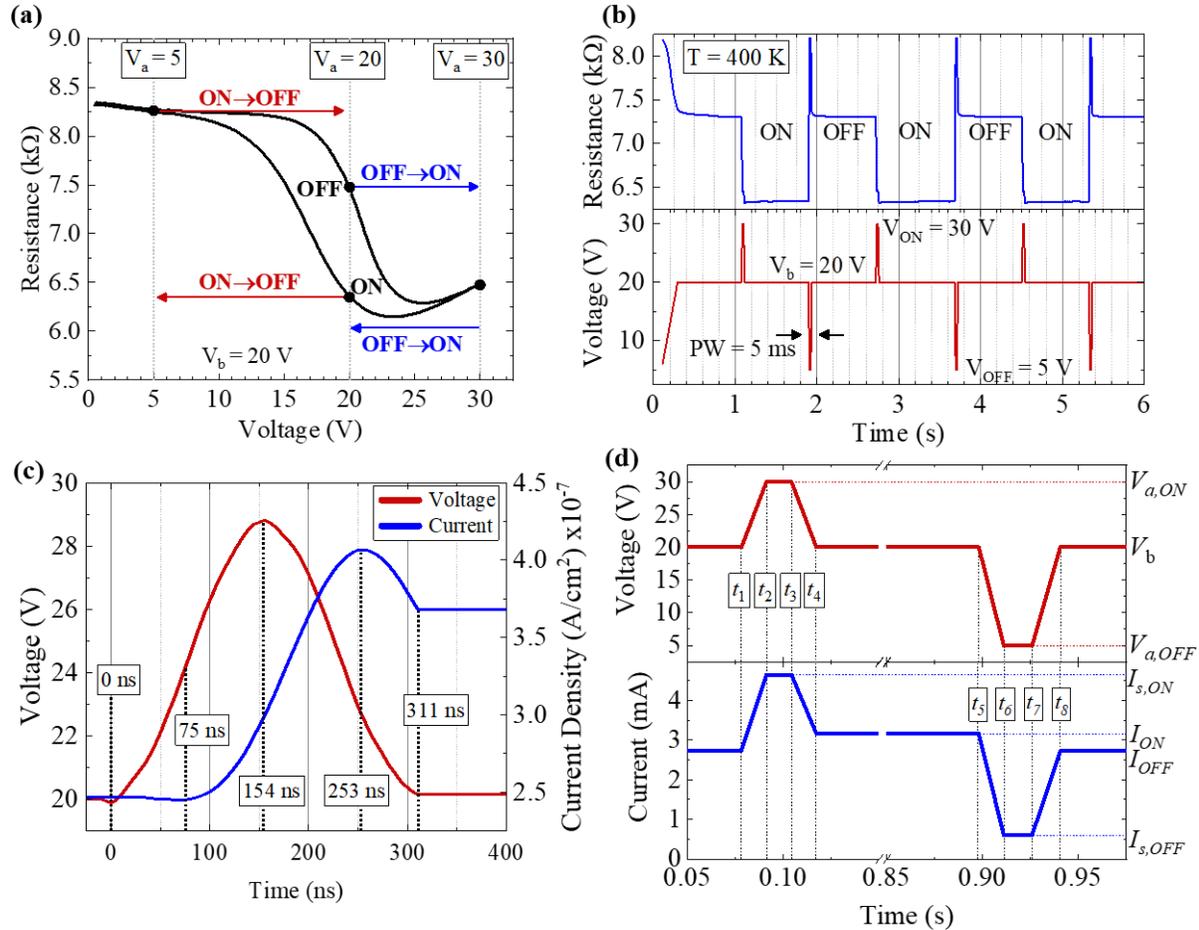

FIG. 5. (a) Pulsed operation state-switching schematic diagram that shows the pulsed voltage amplitudes required to switch between AFM and FM states. At a constant 20 V bias, the FeRh can be in either of the two states, and it will remain in that state assuming it is thermally stabilized. Switching between states can be achieved by applying a short, pulsed voltage of either $V_a = 5$ V (e.g. FM-AFM transition, ON to OFF) or $V_a = 30$ V (e.g. AFM-FM transition, OFF to ON). In (b) we demonstrate this switching capability by showing the transient resistance of the FeRh while pulse biasing the device to switch between ON and OFF states using a $V_b$ of 20 V and $V_a$ of either 5 V (OFF) or 30 V (ON). (c) A high switching speed of 311 ns was observed when switching from OFF to ON. The switching speed shown here was limited by measurement equipment, and should serve as a 'ceiling' for FeRh switching speed capability. (d) Switching parameters used for calculating device power consumption, including $V_{a,ON} = 30$ V, $V_b = 20$ V, $V_{a,OFF} = 5$ V, $I_{s,ON} = 4.646$ mA, $I_{ON} = 3.155$ mA, $I_{OFF} = 2.736$ mA, $I_{s,OFF} = 0.610$ mA, $t_2 - t_1 = 13.045$ ms, $t_3 - t_2 = 13.014$ ms, $t_4 - t_3 = 12.999$ ms, $t_6 - t_5 = 13.053$ ms, $t_7 - t_6 = 14.753$ ms, and $t_8 - t_7 = 14.742$ ms.

To establish an upper bound on the switching speed, we investigated the transient current measured during individual 150 ns voltage pulses (our instrumentation allows for single transient pulses for PW < 5 ms rather than a repeating waveform as shown above for PW > 5 ms). **Figure 5(c)** shows the applied transient voltage (red curve) and measured current (blue curve) during individual pulses where $V_a$ and $V_b$ were 29 V and 20 V, respectively. Within the resolution of our measurement, the device was capable of switching from OFF to ON in 311 ns. During this measurement, the equipment was configured to apply a 150 ns voltage pulse. However, the pulsed voltage had rise/fall times of ~ 150 ns, resulting in a pulse width of more than 300 ns, thus limiting our measurement capability. Therefore, this is an upper bound of the switching time, and we expect actual switching speeds to be much faster. Optical measurements by Pressacco, *et al*. showed the FeRh phase transition occurs on sub-picosecond time scales, suggesting that the operational limit for devices based on the FeRh transition could exceed GHz operating speeds, provided sufficient thermal sinking by the substrate [34]. To understand how this device can be optimized, we evaluate the power switching losses that occur during phase transition. Power consumption was calculated for the ON state and OFF states,

$$P_{ON} = V_b I_{ON}, \tag{2}$$

$$P_{OFF} = V_b I_{OFF}, \tag{3}$$

and also during turn-ON and turn-OFF,

$$P_{s,ON} = \frac{1}{T_{s,ON}} \big[ (t_2 - t_1)(I_{s,ON} - I_{OFF})(V_{a,ON} - V_b) + (t_3 - t_2) I_{s,ON} V_{a,ON}$$
$$+ (t_4 - t_3)(I_{s,ON} - I_{ON})(V_{a,ON} - V_b) \big], \tag{4}$$

$$P_{s,OFF} = \frac{1}{T_{s,OFF}} \big[ (t_6 - t_5)(I_{ON} - I_{s,OFF})(V_b - V_{a,OFF}) + (t_7 - t_6) I_{s,OFF} V_{a,OFF}$$
$$+ (t_8 - t_7)(I_{OFF} - I_{s,OFF})(V_b - V_{a,OFF}) \big] \tag{5}$$

switching cycles [44]. Here, $T_s$ is the switching pulse period, $t_i$ represents the pulsed waveform timings, $I_{ON}$ and $I_{OFF}$ are steady state current amplitudes, $I_{s,ON}$ and $I_{s,OFF}$ are switching current amplitudes resulting from pulsed voltages $V_{a,ON}$ and $V_{a,OFF}$, respectively. The value of each parameter is shown in **Figure 5(d).** Using these parameters, $P_{ON}$, $P_{OFF}$, $P_{s,ON}$, and $P_{s,OFF}$ were found to be 63.11 mW, 54.76 mW, 57.78 mW, and 23.82 mW, respectively.

## IV. DISCUSSION

Based on the switching properties observed in this work, FeRh is comparable to or better than many other candidate phase-change memory materials. Among these, $Ge_2Sb_2Te_5$ is the most commonly used active layer material in phase change memory applications, and has reported read/write times of 150 to 200 ns [45,46]. $TiO_x$ is commonly used for developing memristor networks requiring more complex architectures because it can provide low variability across the area of a chip [47]. Alibart, *et al.* fabricated a $TiO_2$-based memristor that had an $R_{OFF}/R_{ON}$ and write time of 10 and 200 ns, respectively [48]. $HfO_2$ is another commonly used memristor active layer material because of high $R_{OFF}/R_{ON}$ and turn-on slopes of over ten orders of magnitude and 1 mV/decade, respectively [49]. Despite the more intricate architectures of these memory devices, the 311 ns write time demonstrated in this work is comparable in switching speed performance.

Our rudimentary devices will require optimization to maximize $R_{OFF}/R_{ON}$, minimize the relatively high steady-state power consumption ($P_{OFF}$ = 54.763 mW and $P_{ON}$ = 63.105 mW), and ultimately achieve sub-ns switching speeds. There are several strategies for achieving these gains. The $R_{OFF}/R_{ON}$ ratio could be magnified by incorporation into tunneling magnetoresistance type sensors, where the dependence of the tunneling mechanism on the magnetic properties can result in orders of magnitude larger MR changes [50]. The power consumed is dominated by $V_b$ since the pulsed waveform has such a small duty cycle, and the baseline voltage is necessary since it keeps the device temperature near the phase transition. By tuning the width and temperature onset of the FeRh transition, as well as the wire geometry, the required baseline voltage can be drastically reduced. This would also allow for a reduced ON-state power as a smaller pulse would be needed to switch states. The transition temperature of FeRh could also be tuned to lower temperatures by altering the $Fe_xRh_{1-x}$ composition during growth. It was shown previously that the transition temperature can be controlled over a 80 K range by varying the Fe content from $x = 0.40 – 0.49$ or by He-ion implantation [27,51,52]. This would greatly reduce the required $V_b$ and hence the static power consumption. As an example of a potential enhancement in device operating properties, in the current devices, the ON-state power could be reduced by ~32% to 43 mW by lowering $V_b$ to 15 V.

The switching speeds of the devices reported here is comparable to that of conductive filament memristors, which can range from milliseconds to nanoseconds and varies depending on the thickness of the switching medium layer [53–55]. Further reducing the pulse width to the transition speed limit could result in probabilistic switching, allowing for applications of these devices in neuromorphic computing. Existing neuromorphic transistors operate with speeds from kilohertz up to megahertz [6,56,57]. The switching speed of the preliminary device presented here is therefore at least as fast as the state-of-the-art. The 400 K operating temperature used for device proof-of-concept in **Figure 5** was chosen to simulate a realistic operating condition in computing applications. However, in neuromorphic applications, the ideal operating temperature in an advanced system would likely be closer to 310 K [58]. This would have the added benefit of a lower power device. Further, the FeRh device presented here has similar functionality to Mott transistors, which operate based on thermally-induced phase change and can be switched on a sub-ns timescale [59–61]. Similar to other materials used for thermally-induced phase change devices, FeRh is unlikely to experience oxidation effects within the heating/cooling range implemented in this work (*i.e.* 400 K plus the Joule heating contribution). Because there are no stable compounds containing Fe, Rh, and O [62], oxidation is only possible when the Fe-Rh bond is broken. The Fe-Rh bond has a formation energy of -0.055 eV, which is equivalent to the energy provided by a temperature of 638 K. Since the experiments performed in this work were at temperatures less than or equal to 480 K, it is unlikely that any Fe-Rh decomposition would occur. Additionally, based on our findings, it would be possible to fabricate a nonvolatile FeRh device capable of storing memory at zero current which can only be erased by cooling.

## V. CONCLUSIONS

Our comprehensive results establish the feasibility of employing the metamagnetic transition in FeRh as the basis for a very fast, phase-change switch in future computing applications. Joule heating of FeRh wire devices was demonstrated and the geometrical dependencies on the metamagnetic AFM-FM phase transition were investigated. COMSOL simulations and pulsed I-V measurements were used to evaluate the underlying thermal mechanics present during the AFM-FM transition. We demonstrated metamagnetic resistive switching capability with a switching speed of at least 311 ns.

## ACKNOWLEDGMENT

The authors gratefully acknowledge critical assistance from LPS support staff including G. Latini, J. Wood, R. Brun, P. Davis, and D. Crouse.

**APPENDIX A: THERMAL MODEL DESCRIPTION**

**Figure 6(a)** shows the temperature-dependent conductivity functions used to model the heat transfer physics. Here, the temperature-dependent conductivity of FeRh in the AFM and FM states is defined as shown by the blue and green curves, respectively. These functions were obtained by fitting the temperature-dependent conductivity of FeRh in the respective states at temperatures above and below the metamagnetic transition temperature. To transition between the two states, a boundary ordinary differential equation constraint was defined in the model which set a state variable to 1 or 0 corresponding with the AFM and FM states, respectively. A value of 1 causes the domain to follow the AFM temperature-dependent conductivity function while a value of 0 causes it to follow the FM temperature-dependent conductivity function; transition between a state of 0 and 1 is abrupt and solely dependent on the transition temperature of the domain. The transition temperature of each domain was defined by three parameters. The first two parameters are global and define the transition from AFM to FM (430 K) along with the transition from FM to AFM (420 K). The difference between these two parameters gives rise to hysteresis in the metamagnetic transition temperature. Additionally, a random variable is assigned to each FeRh domain, that is seeded by the domain coordinates to keep it fixed for each sampling. Each domain is then sampled from a standard normal distribution, yielding a mean temperature shift of 0 K with ±1 standard deviation corresponding with a ±10 K shift. This random parameter causes each FeRh domain to transition at different temperatures, yielding a gradual change in the total wire resistance that is sampled at the wire terminals. Programmatically, FeRh wire and metal contacts were treated as layered materials, thus accounting for their thickness analytically thereby avoiding the need to mesh them in the out-of-plane direction. Therefore, each mesh element in the FeRh wire region was treated as an individual domain with the maximum domain size ranging from 100 nm to 250 nm.

In the model we chose distinct thermal conductivity and heat capacity parameters for each material. The FeRh thermal conductivity has been recently investigated by Jiméneza et al. using first-principle calculations [63]. It was found that at 400 K the thermal conductivity varies from 31.8 W/m-K to 65.8 W/m-K for antiferromagnetic and ferromagnetic FeRh, respectively. In this work, we have chosen to use a thermal conductivity value of 50 W/m-K since the FeRh device is mainly operated within the AFM-FM transition range. The FeRh heat capacity was set to vary with temperature according to

$$Cp_{FeRh}\left[\frac{J}{\text{kg K}}\right] = 252 + 6.3(T - 150)^{0.95}, \tag{6}$$

which was found by extrapolating data from a report by Cooke *et al.* [64], and is in agreement with additional findings published elsewhere [65]. The thermal conductivity,

$$\kappa_{MgO}\left[\frac{W}{\text{m K}}\right] = \frac{14905\frac{W}{m}}{T} - \left(0.10272\frac{W}{m}\right)T^{0.41267}, \tag{7}$$

and heat capacity,

$$Cp_{MgO}\left[\frac{J}{\text{kg K}}\right] = 47.260 + 5.682\frac{T}{10^3} - 0.873\left(\frac{T}{10^3}\right)^2 + 0.104\left(\frac{T}{10^3}\right)^3 - 1.054\left(\frac{10^3}{T}\right)^2, \tag{8}$$

of the MgO substrate varied with temperature according to previous findings [66]. The thermal conductivity and heat capacity of the gold contacts were 310 W/m-K and 128 J/kg-K, respectively [67].

**Figure 6(b)** shows the resistance of a 10 μm x 100 μm wire under different DC biases as the substrate temperature is increased and serves to verfiy the modeling approach. The simulated wire dimensions and biasing conditions closely match those of **Figure 2(a)**, and indeed the simulation captures the primary behavior observed in experiment. In particular, we observe a hysteresis of approximately 10 – 15 K, a reduction in the onset of transition temperature with DC bias, and an increase in the transition width with DC bias, all of which are reflected in the measured results. The temperature-dependent behavior of the FeRh wire was also found to be highly dependent on the ability of the substrate to dissipate current-induced heating. The inset images in **Figure 6(b)** show volume temperature contour plots within the MgO substrate for the FeRh wire while under a fixed current load of 4 MA/cm$^2$ at different substrate temperatures as labeled. For the 1.0 MA/cm$^2$ condition, the substrate temperature at which the minimum resistance occurs is approximately 440 K. This is approximately 2 standard deviations above the mean transition temperature and correlates to a FM domain transition percentage of >97%. At 2.5 MA/cm$^2$, 4.0 MA/cm$^2$, and 5.0 MA/cm$^2$, the substrate temperature at which the minimum terminal resistance occurs decreases from 415 K to 375 K to 340 K, respectively. Simultaneously, the thermal profile originating at the FeRh wire and extending down into the substrate is clearly apparent. More rapid thermal equilibration within the substrate would lessen the thermal gradient experienced in the wire and concomitantly reduce the amplitude of Joule-heating-induced shift of the minimum resistance temperature. The opposite is also true, whereby a slower thermal equilibration would yield even greater thermal gradients at the expense of slower thermal constants for the wire/substrate system.

In **Figure 6(c)** the simulated wire resistance is plotted as a function of current density for 1-µm-wide wires with lengths of 2.5 µm, 5 µm, 10 µm, and 50 µm. Consistent with the experimental data in **Figure 2(b)**, the simulated wires show a reduction in the current at which the minimum conductivity is observed and a decreased $J_{Cr}$, signifying the onset of the metamagnetic transition. At only 1 µm wide, the shorter wires also begin to display stepwise resistance fluctuations since the transition of individual domains provide greater contributions to the overall resistance. In **Figure 6(c)** a maximum domain size of 100 nm was used to ensure the width of the simulated wire contained at least 10 elements. For **Figure 6(b)** a maximum domain size of 250 nm was used since the wires were 10 µm long and a fixed length of 100 µm. **Figure 6(d)** shows the simulated wire resistance plotted as a function of current density for 100-µm-long wires with widths of 0.3 µm, 1.0 µm, 10 µm, 25 µm, and 50 µm.

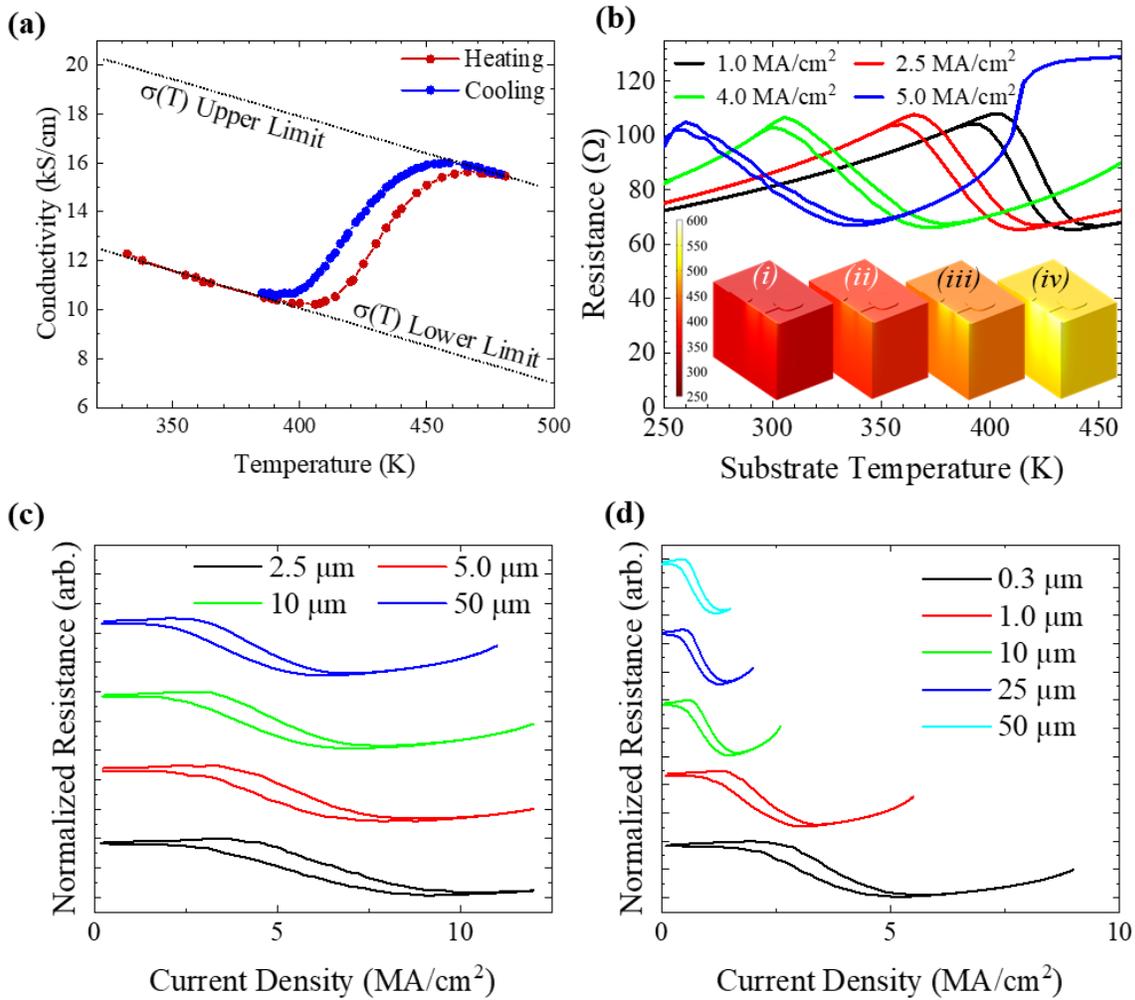

FIG. 6. (a) Experimental conductivity measurements during heating (red), cooling (blue), and linear fits to the conductivity in AFM and FM states shown with dotted lines. (b) Simulated FeRh wire resistance as a function of substrate temperature for varying current densities. Inset shows MgO substrate temperature profiles while applying a current density of 4 MA/cm² at fixed bottom-surface temperatures of (*i*) 350 K, (*ii*) 400 K, (*iii*) 450 K, and (*iv*) 500 K. (c) Simulated FeRh wire resistance as a function of applied current density for a 1-µm-wide wire with varying length. (d) Simulated FeRh wire resistance as a function of applied current density for a 100 µm long wire with varying width. Data shown in (b) and (c) was obtained for a substrate temperature of 400 K.

**APPENDIX B: MODELING OF WIRE TEMPERATURE PROFILE**

Further analysis of the model results are provided in **Figure 7**, including FeRh wire temperature profile within the FeRh wire (x-axis positions ≤ |50| µm) and extended into the Au contacts on either side (x-axis positions ≥ |50| µm) at substrate temperatures of (a) 250 K and (b) 400 K. The temperature profile within the FeRh wire is nearly constant, and referred to as a temperature plateau. Near the metal contacts (within ± 50 µm) the temperature abruptly decreases as the temperature equilibrates with the substrate temperature. This effect is consistent until exceeding a distance of ~ 25 µm from the metal/FeRh interface. Joule heating causes the plateau amplitude to increase and also causes plateau rounding near the metal/FeRh interface. During the thermal cycling simulation, one can envision these profiles as being shifted vertically toward higher temperatures. Then upon reaching the metamagnetic transition temperature, the rate at which this translation occurs begins to slow since the wire resistance and power dissipated by Joule heating ($I^2R$) decreases. This is indeed what we observed when analyzing the first derivative of the terminal resistance with respect to temperature as shown in **Figure 7(c)**. Generally, we see that $dR/dT > 0$ due to the positive temperature coefficient of resistance. However, as the wire transitions from AFM to FM phase, we observe an abrupt decrease in $dR/dT$ such that it is less than 0. Further increasing the substrate temperature, we see $dR/dT > 0$ once the AFM-FM transition is complete. For I = 0.60 mA, the $dR/dT$ transition window is relatively abrupt and the width (T = 395 to 455 K) greatly exceeds the plateau temperature of the wire. Therefore, the $dR/dT$ for this low current sweep is effectively tracing the probability distribution of a domain in the AFM state to transition to the FM state. Upon increasing the current we see the $dR/dT$ minima shift to lower temperatures since less substrate heating is needed to reach the transition. As shown in **Figure 7(c)**, these various trends correlate quite well. Broadening of *dR/dT* transition region is in part caused by the increased curvature of the temperature plateaus near the metal/FeRh interface. Away from the metamagnetic transition, the resistance should vary linearly with temperature, leading to a constant *dR/dT*. However, a slight linear increase is observed in the data. This linear increase is caused by the thermal feedback in the model where an increased substrate temperature causes an increased resistance, which causes increased Joule heating. More Joule heating increases the local wire temperature and further increases the resistance. The increased local resistance is scaled by $I^2$ and therefore increases with current density.

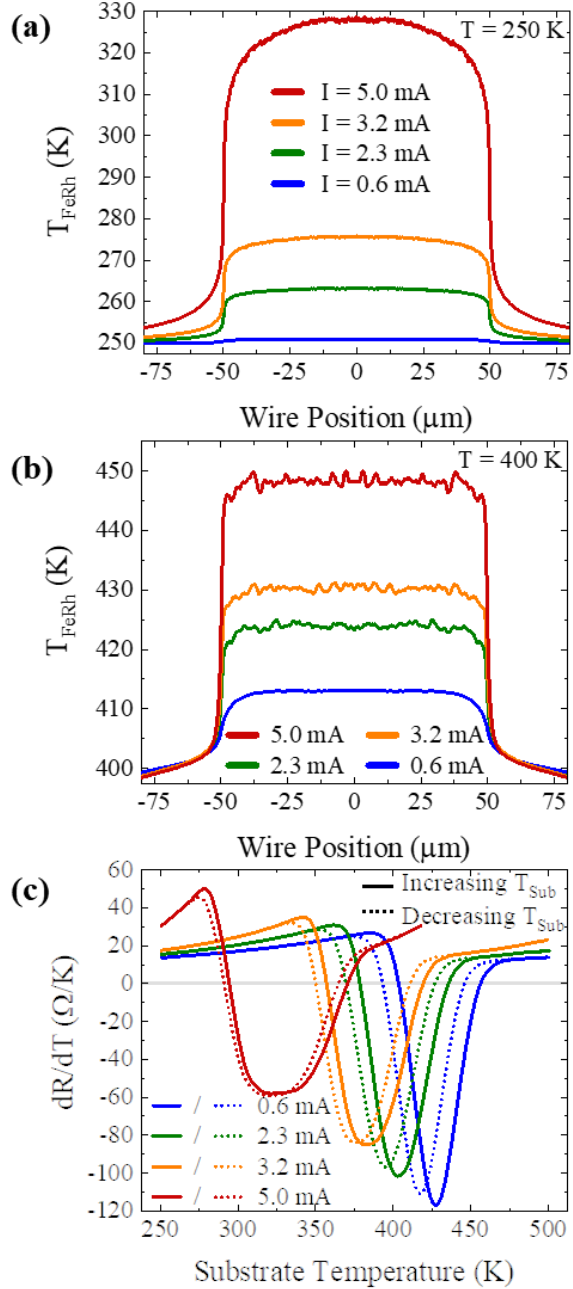

FIG. 7. Position-dependent temperature profile for a substrate temperatures of (a) 250 K and (b) 400 K. The data shown here corresponds to that of **Figure 6(b)** for varying current amplitudes. These findings indicate significant heatsink-effects at the metal/FeRh interface. (c) The first derivative of resistance with respect to temperature, dR/dT, using data from **Figure 6(b)**. Here we observe a 'plateau' region which widens as the current amplitude in increased.

**APPENDIX C: TRANSIENT MODELING**

**Figure 8(a)** shows simulation results where a 30 V voltage pulse of 240 ns pulse width and $V_b$ of 0 V is applied to the wire. Heat dissipation time constants can be extracted from the transient profile using

$$T_{rise} = T_0 + A_f e^{-\frac{x}{\tau_{rf}}} + A_s e^{-\frac{x}{\tau_{rs}}} \tag{9}$$

and

$$T_{decay} = T_0 + B_f e^{-\frac{x}{\tau_{df}}} + B_s e^{-\frac{x}{\tau_{ds}}}, \tag{10}$$

where the time constant abbreviations of *r*, *d*, *f*, and *s* represent rise, decay, fast, and slow, respectively. The time constants were found to be $\tau_{rf} = \tau_{rs} = 200.7\ ns$, $\tau_{df} = 21.8\ ns$, and $\tau_{ds} = 587.2\ ns$. This means that when a 30 V bias is applied with $V_b = 0$ V, the wire temperature is elevated through Joule heating by nearly 40 K over 200 ns. Then, upon removing the 30 V bias, the heat is dissipated from the wire in 587 ns, with approximately 87.5% of the generated heat being dissipated in just 21.8 ns.

Next, we simulate state-switching capability at substrate temperatures and baseline voltages ranging from 275 – 400 K and 12 – 20 volts, respectively, by applying the pulsed waveforms shown in **Figure 8(b)**. The findings are summarized in **Figure 8(c)** and the transient resistance and FeRh wire temperature plots are shown in **Figure 8(d) – (o)**. At temperature of 275 K **(d, e)** and 300 K **(f, g)** the FeRh was unable to switch into the FM state regardless of $V_b$ due to the low substrate temperature. At 325 K **(h, i)** AFM-FM modulation begins at $V_b = 20$ V, but the low substrate temperature allows the FeRh to cool back into the AFM state when no switching bias is applied. State-switching is observed when increasing the substrate temperature to 350 K **(j, k)** and 375 K **(l, m)**, when applying baseline voltages of 18 V and 14 V, respectively. At other biasing conditions the FeRh is either locked into the AFM or FM phase. At 400 K **(n, o)** the FeRh immediately enters the FM phase, but is unable to switch back to AFM as the substrate temperature is too high. **Figure 8(c)** shows substrate temperature versus baseline voltage with AFM, FM, and AFM-FM modulation conditions plotted. In the $T_{Sub}$-$V_b$ regions shaded blue and red, the FeRh was locked in the AFM and FM phases, respectively. However, in-between these two regions we see AFM-FM modulation, and this region of allowed transitions can be described by

$$T(K) = 462.5\ K - (6.25\ K\ V^{-1})V_b \tag{11}$$

which was acquired by fitting as shown in the plot.

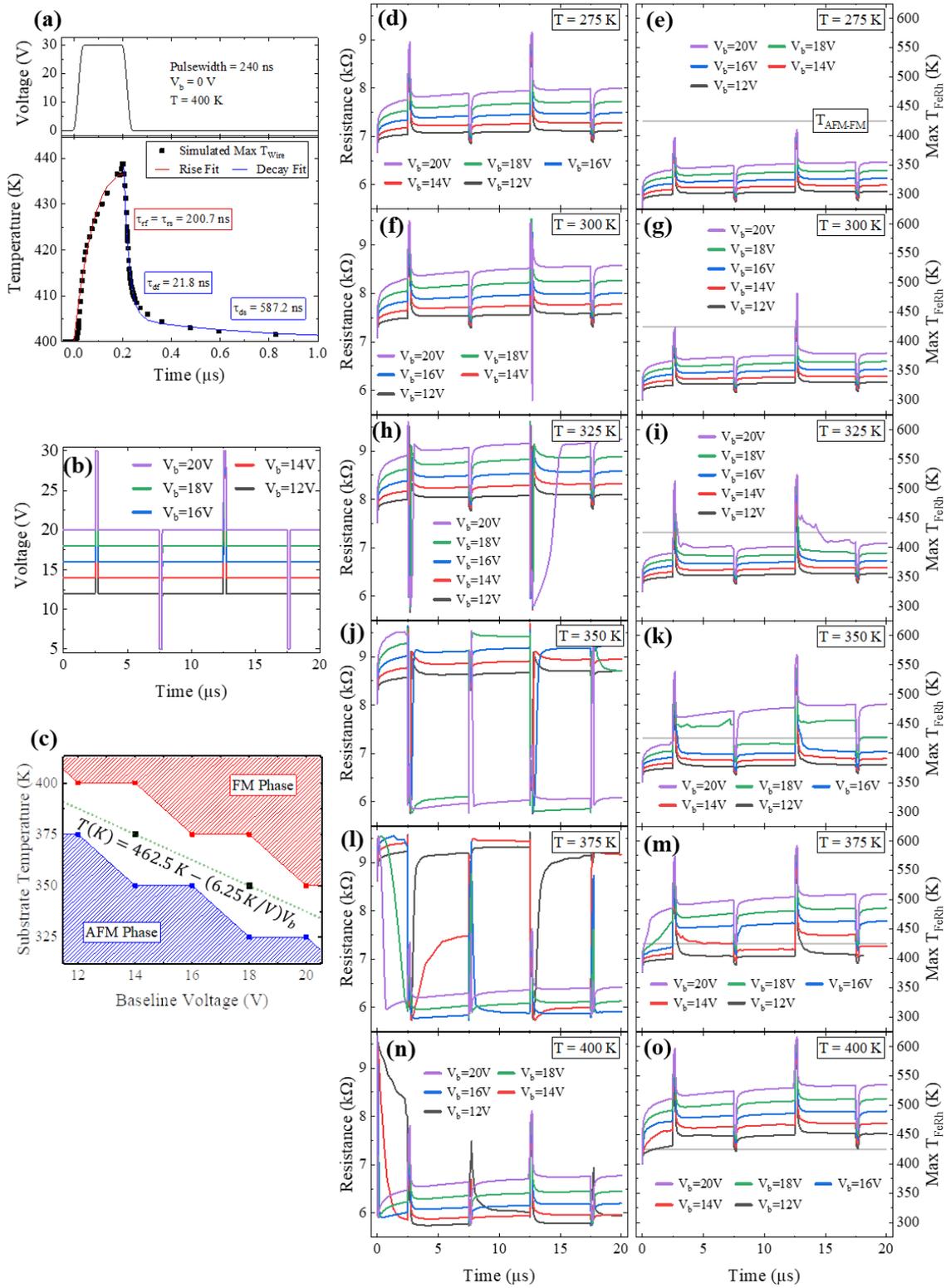

FIG. 8. (a) FeRh wire heat dissipation time constant. (b) Simulated voltage pulse profile. (c) Compiled AFM-FM transition dynamics obtained from the modeled results shown in (d) – (o). The simulated transient data includes FeRh resistance and max wire temperature.

# APPENDIX D: SWITCHING ENDURANCE AND STATE RETENTION

**Figure 9(a)** and **(b)** shows the device switching endurance and state retention, respectively. Endurance was evaluated using the same experimental conditions described for the measurement of **Figure 5(b)**. No noticeable change in the device performance was observed when subject to more than $1\times10^4$ switching cycles. State retention was also investigated by switching into the desired state then monitoring resistance as a function of time for more than one hour. In either case, the resistance reached saturation before the end of the one-hour measurement. $R_{OFF}$ decreased by 24 Ω (0.47%) from the initial value, while $R_{ON}$ increased by 12 Ω (0.24%) from the initial value.

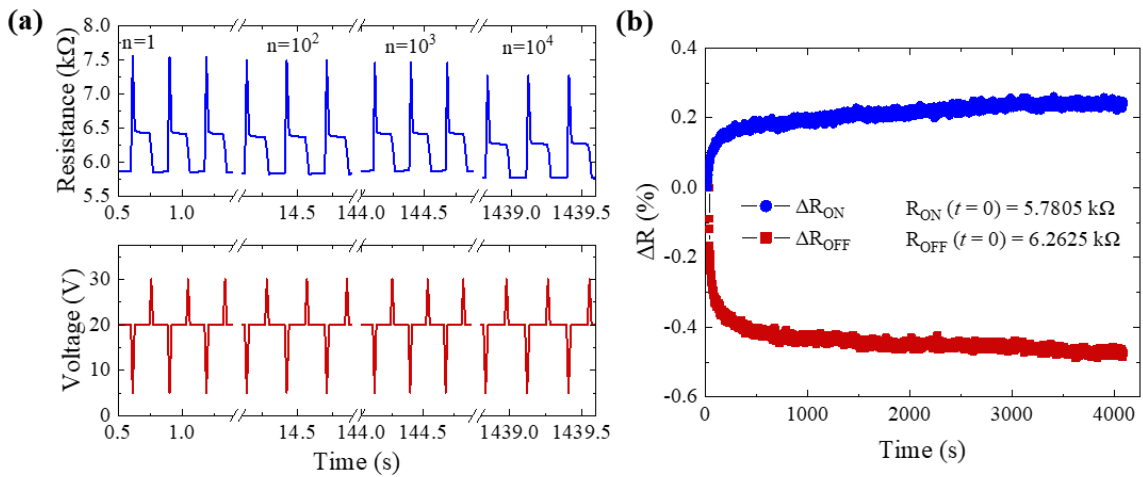

FIG. 9. (a) FeRh endurance was evaluated by switching between the two device states for more than $1\times10^4$ cycles. Only a minimal change in resistance was observed. (b) The device state retention was investigated by switching into the desired state then monitoring transient resistance for more than one hour. The resistance reached saturation before the end of the one-hour measurement. $R_{OFF}$ decreased by 24 Ω (0.47%) from the initial value, while $R_{ON}$ increased by 12 Ω (0.24%) from the initial value.